\documentstyle[12pt,epsf]{article}

\begin{document}

\begin{titlepage}

\hspace*{\fill}\parbox[t]{2.8cm}{DESY 95-040 \\ March 1995}

\vspace*{1cm}

\begin{center}
\large\bf
Equivalence of the Parke-Taylor and the Fadin-Kuraev-Lipatov amplitudes in the
high-energy limit
\end{center}

\vspace*{0.5cm}

\begin{center}
Vittorio Del Duca \\
Deutsches Elektronen-Synchrotron \\
DESY, D-22603 Hamburg , GERMANY
\end{center}

\vspace*{0.5cm}

\begin{center}
\bf Abstract
\end{center}

\noindent

We give a unified description of tree-level multigluon amplitudes in the
high-energy limit. We represent the Parke-Taylor amplitudes and the
Fadin-Kuraev-Lipatov amplitudes in terms of color configurations that are
ordered in rapidity on a two-sided plot. We show that
for the helicity configurations they have in common the
Parke-Taylor amplitudes and the Fadin-Kuraev-Lipatov amplitudes coincide.

\end{titlepage}

\baselineskip=0.8cm

\section{Introduction}

In high-energy hadron collisions multi-jet events are of
phenomelogical interest because they appear as background to top-quark and
electroweak-boson production and to eventual Higgs-boson
production and new-physics signals.
To compute the production rate for multi-jet events, we need to
evaluate amplitudes with multi-parton final states.
These are also of interest {\sl per se} because they yield the radiative
corrections to the total parton cross section, which in the high-energy limit
of perturbative QCD is predicted to have a power-like growth in the parton
center-of-mass energy $\sqrt{\hat s}$ \cite{lip}-\cite{BFKL}.

Multi-parton amplitudes have been computed in the high-energy limit by Fadin,
Kuraev and Lipatov (FKL) \cite{FKL}, who considered the tree-level
production of $n$ gluons in parton-parton scattering in the limit of a
strong rapidity ordering of the produced partons, assuming their
transverse momenta to be all of the same size, $Q$. This kinematic
configuration is termed {\sl multiregge kinematics}. The amplitudes are given
by the exchange of a gluon ladder between the scattering partons
(Fig.\ref{fig:fkl}). FKL made also an ansatz for the leading logarithmic
contribution, in $\ln(\hat s/Q^2)$, of the loop corrections to the
multi-parton amplitudes, to all orders in $\alpha_s$. This changes
the form of the propagators of the gluons exchanged in the $\hat t$ channel,
but preserves the ladder structure of the amplitudes. Using then
$\hat s$-channel unitarity and dispersion relations, FKL computed the total
cross section for a one-gluon ladder exchange and the elastic amplitude for
the exchange of a two-gluon ladder in a color-singlet configuration, i.e. for
the exchange of a perturbative pomeron.

On the other hand, {\sl exact} tree-level amplitudes for the production of $n$
gluons have been computed by Parke and Taylor (PT) \cite{pt} in a helicity
basis, for specific helicity configurations of the incoming
and outgoing gluons. In a helicity basis the color structure of the tree-level
amplitudes may be decomposed as a sum over all the noncyclic permutations
of the gluon color flows.
In a previous work \cite{vd} we have represented the color flows of
the PT amplitudes in terms of color lines in the fundamental representation
of SU($N_c$). Permuting the color flows the color lines appear twisted,
however every configuration may be untwisted
introducing a two-sided plot \cite{bj}. We have shown then that
restricting the PT amplitudes to the multiregge kinematics only
the untwisted configurations with the gluons ordered in rapidity
on the two-sided plot contribute. In ref.\cite{vd} we worked with
the squared PT amplitudes at leading $N_c$, however due to the incoherence of
the leading $N_c$ term in the color sum of the squared PT amplitudes, the
color flows we consider there are the same as the ones of the PT
amplitudes themselves. Thus, also for the PT amplitudes, for which no
approximation in $N_c$ is made, the leading color configurations are the ones
with the gluons ordered in rapidity on the two-sided plot.

In this paper we consider again the sum over the leading color flows
of the PT amplitudes in the multiregge kinematics, and
we show that for the helicity configurations they have in common the PT
amplitudes and the FKL amplitudes are equal.

Besides, we note that the two helicities of each gluon emitted along the
gluon ladder contribute to the same extent to the FKL amplitude,
since changing the helicity of a gluon along the ladder changes the
FKL amplitude only by a phase.

\section{Spinor algebra in multiregge kinematics}
\label{sec:one}

We consider the production of $n+2$ gluons of momentum $p_i$, with
$i=0,...,n+1$ and $n\ge 0$, in the scattering between two gluons of momenta
$p_A$ and $p_B$, and we assume that the produced gluons satisfy
the multiregge kinematics, i.e. we require that the gluons
are strongly ordered in rapidity $y$ and have comparable transverse momentum,
\begin{equation}
y_0 \gg y_1 \gg ...\gg y_{n+1};\qquad |p_{i\perp}|\simeq|p_{\perp}|\,
.\label{mreg}
\end{equation}
The Mandelstam invariants (\ref{inv}) (Appendix A) take then the approximate
form,
\begin{eqnarray}
\hat s &\simeq& |p_{0\perp}| |p_{n+1\perp}| e^{y_0-y_{n+1}}\, ,\nonumber \\
\hat s_{Ai} &\simeq& - |p_{0\perp}| |p_{i\perp}| e^{y_0-y_i}\, ,\label{invb}\\
\hat s_{Bi} &\simeq& - |p_{i\perp}| |p_{n+1\perp}| e^{y_i-y_{n+1}}\, ,\nonumber
\\ \hat s_{ij} &\simeq& |p_{i\perp}| |p_{j\perp}| e^{|y_i-y_j|}\, .\nonumber
\end{eqnarray}

In the calculation of helicity amplitudes, polarization vectors are
expressed in terms of massless spinors. Thus, we recall here and in Appendix B
concepts and notations of spinor algebra.
Massless Dirac spinors $\psi_{\pm}(p)$ of fixed helicity are
defined by the projection,
\begin{equation}
\psi_{\pm}(p) = {1\pm \gamma_5\over 2} \psi(p)\, .\label{spi}
\end{equation}
We use for the spinors the shorthand notation of ref.\cite{cina},
\begin{eqnarray}
\psi_{\pm}(p) &=& |p\pm\rangle, \qquad \overline{\psi_{\pm}(p)} =
\langle p\pm|\, ,\nonumber\\
\langle p k\rangle &=& \langle p- | k+ \rangle = \overline{\psi_-(p)}
\psi_+(k)\, ,\label{cpro}\\
\left[pk\right] &=& \langle p+ | k- \rangle = \overline{\psi_+(p)}\psi_-(k)\,
.\nonumber
\end{eqnarray}
In the multiregge kinematics the spinor products (\ref{spro}), computed in
Appendix B, reduce to
\begin{eqnarray}
\langle p_i p_j\rangle &\simeq& -\sqrt{|p_{i\perp}|\over |p_{j\perp}|}\,
p_{j\perp} \exp({y_i-y_j\over 2})\, \qquad {\rm for}\, y_i>y_j ,\nonumber\\
\langle p_A p_i\rangle &\simeq& -\sqrt{|p_{0\perp}|\over |p_{i\perp}|}\,
p_{i\perp} \exp({y_0-y_i\over 2})\, ,\label{ypro}\\ \langle p_i p_B\rangle
&\simeq& -\sqrt{|p_{i\perp}||p_{n+1\perp}|}\, \exp({y_i-y_{n+1}\over 2})\,
,\nonumber\\ \langle p_A p_B\rangle &\simeq& -\sqrt{|p_{0\perp}|
|p_{n+1\perp}|} \exp({y_0-y_{n+1}\over 2})\, ,\nonumber
\end{eqnarray}
where we have used the complex notation (\ref{com}) and we have expressed
$p_A^+$ and $p_B^-$ through the momentum conservation
(\ref{nkin}) (Appendix A) and kept only the leading contribution.

\section{Parke-Taylor amplitudes in multiregge kinematics}
\label{sec:two}

After having set up the kinematics we are going to use in this and in
the next section, we introduce the PT amplitudes, i.e. {\sl exact} tree-level
amplitudes for the production of $n$ gluons in a specific helicity
configuration. Sorting out the color structure, a tree-level multigluon
amplitude in a helicity basis may be written as \cite{mp}
\begin{equation}
M_n = \sum_{[A,0,...,n+1,B]} {\rm tr}(\lambda^a\lambda^{d_0} \cdots
\lambda^{d_{n+1}} \lambda^b) \, m(\tilde{p}_A,\epsilon_A; p_0,\epsilon_0;...;
p_{n+1},\epsilon_{n+1};\tilde{p}_B,\epsilon_B)\, ,\label{one}
\end{equation}
where $a,d_0,..., d_{n+1},b$, and $\epsilon_A,
\epsilon_0,..., \epsilon_B$ are respectively the colors and the
polarizations of the gluons, the $\lambda$'s are the color matrices in the
fundamental representation of SU($N_c$) and the sum is over the noncyclic
permutations of the set $[A,0,...,B]$. The gauge-invariant subamplitudes,
$m(\tilde{p}_A,\epsilon_A; p_0,\epsilon_0;...; p_{n+1},\epsilon_{n+1};
\tilde{p}_B,\epsilon_B)$, are known for a few specific helicity configurations
\cite{pt}, \cite{kos}. Considering all the momenta as outgoing, the PT
amplitudes describe the {\sl maximally helicity-violating}
configurations $(-,-,+,...,+)$ \cite{pt}, \cite{mp}\footnote{Note that
eq.(\ref{two}) differs for the $\sqrt{2}$ factor from the expression
given in ref.\cite{mp}, because we use the standard normalization of
the $\lambda$ matrices, ${\rm tr}(\lambda^a\lambda^b) = \delta^{ab}/2$.},
\begin {equation}
i m(-,-,+,...,+) = i\, 2^{2+n/2}\, g^{n+2}\, {\langle p_i p_j\rangle^4\over
\langle \tilde{p}_A p_0\rangle \cdots\langle p_{n+1} \tilde{p}_B\rangle
\langle \tilde{p}_B \tilde{p}_A\rangle}\, ,\label{two}
\end{equation}
where $i$ and $j$ are the gluons of negative helicity. The configurations
$(+,+,-,...,-)$ are then obtained by replacing the $\langle p k\rangle$
products with $\left[k p\right]$ products. In eq.(\ref{two})
the following representation for the gluon polarization vector has been used
\cite{mp},
\begin {equation}
\epsilon_{\mu}^{\pm}(p,k) = \pm {\langle p\pm |\gamma_{\mu}| k\pm\rangle\over
\sqrt{2} \langle k\mp | p\pm \rangle}\, ,\label{hpol}
\end{equation}
which enjoys the properties
\begin {eqnarray}
\epsilon_{\mu}^{\pm *}(p,k) &=& \epsilon_{\mu}^{\mp}(p,k)\, ,\label{pola}\\
\epsilon_{\mu}^{\pm}(p,k)\cdot p &=& \epsilon_{\mu}^{\pm}(p,k)\cdot k = 0\,
,\label{polc}\\
\sum_{\nu=\pm} \epsilon_{\mu}^{\nu}(p,k) \epsilon_{\rho}^{\nu *}(p,k) &=&
- g_{\mu\rho} + {p_{\mu} k_{\rho} + p_{\rho} k_{\mu}\over p\cdot k}\,
,\label{polb}
\end{eqnarray}
where $k$ is an arbitrary light-like momentum. The sum (\ref{polb}) is
equivalent to use an axial, or physical, gauge.

{}From the spinor
products (\ref{ypro}), we see that in the multiregge kinematics the PT
amplitudes (\ref{two}) for which the numerator is the largest are the ones for
which the pair of negative-helicity gluons is one of the following,
\begin {equation}
(A,B),\quad (A,n+1),\quad (B,0),\quad (0,n+1)\, .\label{neg}
\end{equation}
%
We focus on the first pair, and fix $\tilde{p}_A = -p_A$ and $\tilde{p}_B =
-p_B$. From eq.(\ref{two}) we have,
\begin {equation}
i m(-p_A,-; p_0,+;...; p_{n+1},+; -p_B,-) = i\, 2^{2+n/2}\, g^{n+2}\,
{\langle p_A p_B\rangle^4\over \langle p_A p_0\rangle \cdots\langle p_{n+1}
p_B\rangle \langle p_B p_A\rangle}\, ,\label{twob}
\end{equation}
where the gluons $p_i$, with $i=1,...,n$, are all emitted with
helicity $\nu = +$.

\begin{figure}[htb]
\vspace{12pt}
\vskip 0cm
\epsfysize=6cm
\centerline{\epsffile{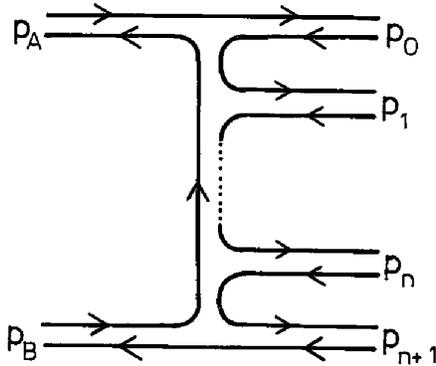}}
\vspace{18pt}
\vskip-1cm
\caption{PT amplitude with color ordering [$A,0,...,n+1,B$].}
\label{fig:one}
\vspace{12pt}
\end{figure}

We put then eq.(\ref{twob}) back into eq.(\ref{one}) and examine in detail
all the color orderings, as we did in ref.\cite{vd}, but here we do it at
the amplitude level. We start with the ordering [$A,0,...,n+1,B$]
(Fig.\ref{fig:one}). Using the spinor products (\ref{ypro}) and the first
of the identities (\ref{flip}) (Appendix B), the
string of spinor products in the denominator of eq.(\ref{twob}) is
\begin {equation}
\langle p_A p_0\rangle \cdots\langle p_{n+1} p_B\rangle \langle p_B p_A\rangle
\simeq (-1)^{n+1} \langle p_A p_B\rangle^2 \prod_{i=0}^{n+1} p_{i\perp}\,
.\label{tre}
\end{equation}
It is easy to see by explicit calculation that every other color
configuration, for which we keep fixed the position of gluons $A$ and $B$
in the color ordering and permute the outgoing gluons,
gives a larger contribution to eq.(\ref{tre})
and so a subleading contribution, of ${\cal O}(e^{-|y_i-y_j|})$, to
eq.(\ref{one}). We note that untwisting the color lines on a configuration
with permuted outgoing gluons, the color ordering we obtain is different from
the rapidity ordering. Thus the leading color configuration in multiregge
kinematics is the one whose
untwisted lines respect the rapidity ordering (Fig.\ref{fig:one}).

\begin{figure}[htb]
\vspace{12pt}
\vskip 0cm
\epsfysize=6cm
\centerline{\epsffile{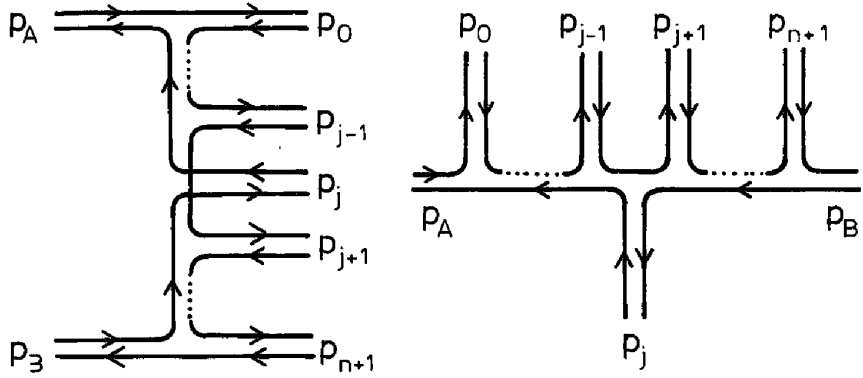}}
\vspace{18pt}
\vskip-1cm
\caption{$a)$ PT amplitude with color ordering [$A,0,...,j-1,j+1,...,n+1,B,
j$], and $b)$ its untwisted version on the two-sided plot.}
\label{fig:two}
\vspace{12pt}
\end{figure}

Next, we move gluon $B$ one step to the left and consider the color orderings
[$A,0,...,j-1,j+1,...,n+1,B,j$], with
$j=0,...,n+1$ (Fig.\ref{fig:two}a). Untwisting the color lines, we get
gluon $j$ on the back of the plot (Fig.\ref{fig:two}b). We
compute then the string of spinor products,
\begin {equation}
\langle p_A p_0\rangle \cdots\langle p_{j-1} p_{j+1}\rangle\cdots \langle
p_{n+1} p_B\rangle \langle p_B p_j\rangle \langle p_j p_A\rangle
\simeq (-1)^n \langle p_A p_B\rangle^2 \prod_{i=0}^{n+1} p_{i\perp}\,
.\label{four}
\end{equation}
We note that the result is independent of which gluon we have taken to the
back of the plot in Fig.\ref{fig:two}b. As compared to the string of spinor
products in eq.(\ref{tre}) we have one more product with reversed order
of the momenta; the first of the identities
(\ref{flip}) (Appendix B) then entails one more minus sign in eq.(\ref{four}).
Next, we note that every permutation of the gluons on the front of the
plot of Fig.\ref{fig:two}b gives a larger contribution
to eq.(\ref{four}) and so a subleading contribution, of ${\cal O}
(e^{-|y_i-y_j|})$, to eq.(\ref{one}). Thus the leading color configurations
are the ($n+2$) configurations whose untwisted lines respect the
rapidity ordering on the front of the plot of Fig.\ref{fig:two}b.

\begin{figure}[htb]
\vspace{12pt}
\vskip 0cm
\epsfysize=6cm
\centerline{\epsffile{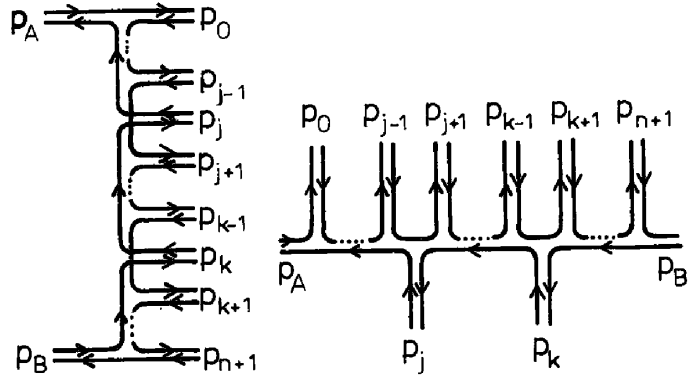}}
\vspace{18pt}
\vskip-0.5cm
\caption{$a)$ PT amplitude with color ordering [$A,0,...,j-1,j+1,...,k-1,
k+1,...,n+1,B,k,j$], and $b)$ its untwisted version on the two-sided plot.}
\label{fig:tre}
\vspace{12pt}
\end{figure}

Then we move gluon $B$ further to the left and consider the color orderings
[$A,0,...,j-1,j+1,...,k-1,k+1,...,n+1,B,k,j$], with $j,k=0,...,n+1$ and
$j < k$ (Fig.\ref{fig:tre}a). Untwisting the color lines, we get
gluons $j$ and $k$ on the back of the plot (Fig.\ref{fig:tre}b).
The related string of spinor products is,
\begin{eqnarray}
& &\langle p_A p_0\rangle \cdots\langle p_{j-1} p_{j+1}\rangle\cdots \langle
p_{k-1} p_{k+1}\rangle\cdots \langle p_{n+1} p_B\rangle \langle p_B p_k\rangle
\langle p_k p_j\rangle \langle p_j p_A\rangle \simeq \nonumber\\
& & (-1)^{n+1} \langle p_A p_B\rangle^2 \prod_{i=0}^{n+1} p_{i\perp}\,
.\label{five}
\end{eqnarray}
The same considerations we have done after eq.(\ref{four}) apply to this
configuration. We just note more that every permutation
of the gluons on the front of the plot {\sl or} on the back of the plot
yields a larger contribution to eq.(\ref{five}) and so a subleading
contribution, of ${\cal O}(e^{-|y_i-y_j|})$, to eq.(\ref{one}). Thus the
leading color configurations are the
$\left(\begin{array}{c} n+2\\ 2\end{array}\right)$ configurations whose
untwisted lines respect the rapidity ordering on both the sides of the plot
of Fig.\ref{fig:tre}b.

We can then proceed further by taking gluon $B$ one more step to the left,
i.e. by considering three gluons on the back of the plot, and so on.
Taking gluon $B$ all the way to the left, we will have
exhausted all the $(n+3)!$ noncyclic permutations of the color ordering
$[A,0,...,B]$.
Substituting then eq.(\ref{twob}), (\ref{tre}), (\ref{four}) and (\ref{five})
into eq.(\ref{one}), we obtain
\begin{eqnarray}
& & i M(-p_A,-; p_0,+;...; p_{n+1},+; -p_B,-) \simeq i\, (-1)^{n+1}\,
2^{2+n/2}\, g^{n+2}\, {\hat s}\, {1\over \prod_{i=0}^{n+1} p_{i\perp}}
\nonumber\\ & & \times {\rm tr}\left(\lambda^a\lambda^{d_0} \cdots
\lambda^{d_{n+1}} \lambda^b -\sum_{j=0}^{n+1} \lambda^a\lambda^{d_0} \cdots
\lambda^{d_{j-1}} \lambda^{d_{j+1}}\cdots \lambda^{d_{n+1}} \lambda^b
\lambda^{d_j}\right. \label{ypt}\\ & & \left. + \sum_{j<k} \lambda^a
\lambda^{d_0} \cdots \lambda^{d_{j-1}} \lambda^{d_{j+1}}\cdots
\lambda^{d_{k-1}} \lambda^{d_{k+1}}\cdots\lambda^{d_{n+1}} \lambda^b
\lambda^{d_k} \lambda^{d_j} + \cdots\right)\, ,\nonumber
\end{eqnarray}
where the color orderings which contribute to eq.(\ref{ypt}) in the
multiregge kinematics are given by the $2^{n+2}$ configurations which respect
the rapidity ordering on the two-sided plot. Using nested commutators (cf.
eq.(\ref{fpro})), eq.(\ref{ypt}) may be written as,
\begin{eqnarray}
& & i M(-p_A,-; p_0,+;...; p_{n+1},+; -p_B,-) \simeq \nonumber\\ & & i\,
(-1)^{n+1}\, 2^{2+n/2}\, g^{n+2}\, {\hat s}\, {1\over \prod_{i=0}^{n+1}
p_{i\perp}}\, {\rm tr}\left(\lambda^a\,\left[\lambda^{d_0},\left[
\lambda^{d_1},...,\left[\lambda^{d_{n+1}}, \lambda^b\right]\right]\right]
\right)\, .\label{mrpt}
\end{eqnarray}

As noted after eq.(\ref{two}), the configurations $(+,+,-,...,-)$ are obtained
by replacing the $\langle p k\rangle$ products with $\left[k p\right]$
products. Because of eq.(\ref{flip}) (Appendix B) this amounts to exchange
$\prod_i p_{i\perp}$ with $\prod_i p_{i\perp}^*$ in eq.(\ref{mrpt}).

Finally, for the other helicity configurations of eq.(\ref{neg}) we obtain,
from eq.(\ref{two}) and the spinor products (\ref{ypro}),
\begin{eqnarray}
M(-p_A,+; p_0,-;...; p_{n+1},+; -p_B,-) &=& M(-p_A,-; p_0,+;...; p_{n+1},+;
-p_B,-)\, ,\nonumber\\
M(-p_A,-; p_0,+;...; p_{n+1},-; -p_B,+) &=& M(-p_A,+; p_0,-;...; p_{n+1},-;
-p_B,+)\, ,\label{othel}\\ &=& \left({p_{n+1\perp}\over p_{n+1\perp}^*}
\right)^2\, M(-p_A,-; p_0,+;...; p_{n+1},+; -p_B,-)\, ,\nonumber
\end{eqnarray}
where the gluons $p_i$, with $i=1,...,n$, are as usual all emitted with
helicity $\nu = +$.

\section{The Fadin-Kureav-Lipatov amplitudes at fixed helicities}
\label{sec:fkl}

\begin{figure}[htb]
\vspace*{-4.5cm}
\hspace*{1.5cm}
\epsfxsize=15cm \epsfbox{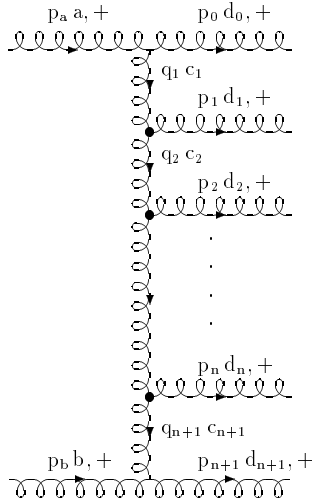}
\vspace*{-10.cm}
\caption{FKL amplitude for fixed gluon helicities. The blobs remind that
Lipatov vertices are used for the gluon emissions along the ladder.}
\label{fig:fkl}
\end{figure}

The tree-level amplitude for the production of $n+2$ gluons in the
multiregge kinematics has been computed in ref.\cite{FKL}
(Fig.\ref{fig:fkl}), and it is
\begin{eqnarray}
iM^{ad_0...d_{n+1}b}_{\nu_A\nu_0...\nu_{n+1}\nu_B} &\simeq& 2i\, {\hat s}
\left(i g\, f^{ad_0c_1}\, \Gamma^{\mu_A\,\mu_0}\right)\,
\epsilon_{\mu_A}^{\nu_A*}(p_A) \epsilon_{\mu_0}^{\nu_0}(p_0)\, {1\over\hat t_1}
\nonumber\\ &\cdot& \left(i g\, f^{c_1d_1c_2}\, C^{\mu_1}(q_1,q_2)\right)\,
\epsilon_{\mu_1}^{\nu_1}(p_1)\, {1\over \hat t_2} \nonumber\\ &\cdot&
\label{ntree}\\ &\cdot&\nonumber\\ &\cdot& \left(i g\, f^{c_nd_nc_{n+1}}\,
C^{\mu_n}(q_n,q_{n+1})\right)\, \epsilon_{\mu_n}^{\nu_n}(p_n)\,
{1\over \hat t_{n+1}} \nonumber\\ &\cdot& \left(i g\, f^{bd_{n+1}c_{n+1}}\,
\Gamma^{\mu_b\,\mu_{n+1}}\right)\, \epsilon_{\mu_B}^{\nu_B*}(p_B)
\epsilon_{\mu_{n+1}}^{\nu_{n+1}}(p_{n+1})\, ,\nonumber
\end{eqnarray}
where the $\nu$'s are the helicities, the $q$'s are the momenta of the gluons
exchanged in the $\hat t$ channel, and $\hat t_i = q_i^2 \simeq
-|q_{i\perp}|^2$. The $\Gamma$-tensors are helicity-conserving tensors,
\begin{eqnarray}
\Gamma^{\mu_A\mu_0} &=& g^{\mu_A\mu_0} -
{p_A^{\mu_0} p_B^{\mu_A} + p_B^{\mu_0} p_0^{\mu_A} \over p_A\cdot p_B} -
\hat{t}_1\, {p_B^{\mu_0} p_B^{\mu_A}\over 2 (p_A\cdot p_B)^2}\, ,\label{gamm}\\
\Gamma^{\mu_B\mu_{n+1}} &=& g^{\mu_B\mu_{n+1}} -
{p_A^{\mu_B} p_B^{\mu_{n+1}} + p_A^{\mu_{n+1}} p_{n+1}^{\mu_B} \over p_A\cdot
p_B} - \hat{t}_{n+1}\, {p_A^{\mu_{n+1}} p_A^{\mu_B}\over 2 (p_A\cdot p_B)^2}\,
,\nonumber
\end{eqnarray}
and the Lipatov vertex \cite{lip} is
\begin{equation}
C^{\mu}(q_i,q_{i+1}) = \left[(q_i+q_{i+1})^{\mu}_{\perp}\, -\,
\left({\hat s_{Ai}\over\hat s}\,+\,2{\hat t_{i+1}\over\hat s_{Bi}}\right)
p_B^{\mu}\, + \left({\hat s_{Bi}\over\hat s}\,+\,2{\hat t_i\over\hat s_{Ai}}
\right) p_A^{\mu}\right]\, ,\label{nver}
\end{equation}
with $q_{i\perp}^{\mu}=(0,0;q_{i\perp})$ and with the Mandelstam invariants as
given in eq.(\ref{invb}). The $\Gamma$-tensors and the Lipatov vertex are
gauge invariant.


Helicity conservation at the production vertices for the first and the
last gluon along the ladder in
eq.(\ref{ntree}) yields the four helicity configurations (\ref{neg}).
For sake of comparison with the PT amplitudes we choose the helicity
configuration of eq.(\ref{twob}). We invert then the momenta of gluons $A$
and $B$ and we obtain for the FKL amplitude the configuration of
Fig.\ref{fig:fkl}. There is however no restriction in eq.(\ref{ntree}) on
the helicities of the gluons produced from the Lipatov vertices along the
ladder.

We choose the representation (\ref{hpol}) for the polarizations. As noted
after eq.(\ref{polb}), this is equivalent to use a physical gauge. Then
we must specify a reference vector with respect to which
we compute the polarization vectors in eq.(\ref{ntree}), but thanks to gauge
invariance the choice is arbitrary. For the polarization
of gluons $p_A$ and $p_0$ we choose $p_B$ as reference vector, while for
the polarization of gluons $p_B$ and $p_i$ with $i=1,...,n+1$ we choose $p_A$.
Following the nomenclature of ref.\cite{lipat} we call the former a
{\sl right} gauge (R) and the latter a {\sl left} gauge (L).

In order to facilitate the contraction with the Lipatov vertex
in L or R gauges it is convenient to decompose a polarization
vector in light-cone or Sudakov components. Using then the property
(\ref{polc}) we obtain \cite{lipat},
\begin{eqnarray}
\epsilon_L^{\mu}(p) &=& \epsilon_{L\perp}^{\mu} - {p\cdot\epsilon_{L\perp}\over
p\cdot p_A} p_A^{\mu}\, ,\label{lhe}\\
\epsilon_R^{\mu}(p) &=& \epsilon_{R\perp}^{\mu} - {p\cdot\epsilon_{R\perp}\over
p\cdot p_B} p_B^{\mu}\, ,\nonumber
\end{eqnarray}
with $\epsilon_{L,R\perp}^{\mu} =(0,0;\epsilon_{L,R\perp})$, and
$\epsilon_{L,R}^2 = \epsilon_{L,R\perp}^2 = -1$.

Contracting the polarization vector in the L gauge (\ref{lhe}) with the
Lipatov vertex (\ref{nver}), we have \cite{lipat}, \cite{mark}
\begin{equation}
\epsilon_L(p_i)\cdot C(q_i,q_{i+1}) = 2\, |q_{i+1\perp}|^2\, \left(
{q_{i+1\perp}^{\mu}\over |q_{i+1\perp}|^2} + {p_{i\perp}^{\mu}\over
|p_{i\perp}|^2}\right)\, \epsilon_{L\perp}^{\mu}\, .\label{vera}
\end{equation}
Using eq.(\ref{staa}) (Appendix C) and the complex notation (\ref{com})
(Appendix A) we may
write the contraction (\ref{vera}) for the helicity $\nu=+$ as
\begin{equation}
\epsilon_L^+(p_i)\cdot C(q_i,q_{i+1}) = -\sqrt{2}\, {q_{i\perp}^* q_{i+1\perp}
\over p_{i\perp}^*}\, ,\label{verb}
\end{equation}
from which, using the conversion table (\ref{conv}) (Appendix C) between the
representations (\ref{hpol}) and (\ref{lhe}), the contraction of the
Lipatov vertex with the gluon polarization in eq.(\ref{ntree}) is,
\begin{equation}
\epsilon^+(p_i, p_A)\cdot C(q_i,q_{i+1}) = \sqrt{2}\, {q_{i\perp}^*
q_{i+1\perp}\over p_{i\perp}}\, .\label{verc}
\end{equation}
{}From eq.(\ref{exh}) (Appendix C), the contractions of the helicity-conserving
tensors (\ref{gamm}) with the gluon polarizations are\footnote{Note that we
obtain the result of eq.(\ref{contra}) also by using the simpler
helicity-conserving tensor $g^{\mu\nu}$, however $g^{\mu\nu}$ is not
gauge invariant as it can be seen
by using it and changing reference vectors in eq.(\ref{contra}).},
\begin{eqnarray}
\Gamma^{\mu_B\mu_{n+1}}\,\epsilon^{+*}_{\mu_B}(p_B, p_A)\,
\epsilon^+_{\mu_{n+1}}(p_{n+1}, p_A) &=& -{p_{n+1\perp}^*\over
p_{n+1\perp}}\, ,\label{contra}\\
\Gamma^{\mu_A\mu_0}\,\epsilon^{+*}_{\mu_A}(p_A, p_B)\, \epsilon^+_{\mu_0}(p_0,
p_B) &=& -1\, .\nonumber
\end{eqnarray}
Substituting eq.(\ref{verc}) and (\ref{contra}) into eq.(\ref{ntree}), the FKL
amplitude in the helicity configuration of eq.(\ref{twob}) becomes
\begin{eqnarray}
& & i M(p_A,+; p_0,+;...; p_{n+1},+; p_B,+) = \label{fklh}\\ & & 2i\, (-1)^n\,
2^{n/2}\, (ig)^{n+2}\, {\hat s}\, {1\over \prod_{i=0}^{n+1} p_{i\perp}}\,
f^{ad_0c_1}\, f^{c_1d_1c_2}\,\cdots f^{c_nd_nc_{n+1}}\,
f^{bd_{n+1}c_{n+1}}\, .\nonumber
\end{eqnarray}

Next, we need to convert the product of structure constants into the trace
of a product of $\lambda$-matrices. From the algebra of the $\lambda$-matrices
we have
\begin{eqnarray}
& & f^{abc} = -2i\, {\rm tr}\left([\lambda^a,\lambda^b]\,\lambda^c\right)\,
,\label{alg}\\ & & \sum_a \lambda^a_{ij} \lambda^a_{kl} = {1\over 2} \left[
\delta_{il} \delta_{jk} - {1\over N_c} \delta_{ij} \delta_{kl}\right]\,
,\nonumber
\end{eqnarray}
from which, using the antisymmetry of the structure constants and the
ciclicity of the trace, we have
\begin{equation}
f^{abz}\, {\rm tr}\left(\lambda^z,\left[\lambda^c,\left[\lambda^d,...,\left[
\lambda^x,\lambda^y\right]\right]\right]\right) = -i\, {\rm tr}\left(\lambda^a
\left[\lambda^b,\left[\lambda^c,\left[\lambda^d,...,\left[\lambda^x,\lambda^y
\right]\right]\right]\right]\right)\, .\label{nalg}
\end{equation}
Using eq.(\ref{alg}) and (\ref{nalg}), it is then easy to see that,
\begin{eqnarray}
& & f^{ad_0c_1}\, f^{c_1d_1c_2}\,\cdots f^{c_nd_nc_{n+1}} f^{bd_{n+1}c_{n+1}} =
\nonumber\\ & & -2\, (-i)^{n+2}\, {\rm tr}\left(\lambda^a\,\left[\lambda^{d_0},
\left[\lambda^{d_1},...,\left[\lambda^{d_{n+1}}, \lambda^b\right]\right]\right]
\right) = \label{fpro}\\ & &
-2\, (-i)^{n+2}\, {\rm tr}\left(\lambda^a\lambda^{d_0} \cdots
\lambda^{d_{n+1}} \lambda^b -\sum_{j=0}^{n+1} \lambda^a\lambda^{d_0} \cdots
\lambda^{d_{j-1}} \lambda^{d_{j+1}}\cdots \lambda^{d_{n+1}} \lambda^b
\lambda^{d_j}\right. \nonumber\\ & & \left. + \sum_{j<k} \lambda^a
\lambda^{d_0} \cdots \lambda^{d_{j-1}} \lambda^{d_{j+1}}\cdots
\lambda^{d_{k-1}} \lambda^{d_{k+1}}\cdots\lambda^{d_{n+1}} \lambda^b
\lambda^{d_k} \lambda^{d_j} + \cdots\right)\, ,\nonumber
\end{eqnarray}
which shows that also for the FKL amplitudes the only configurations
which contribute are the $2^{n+2}$ color
configurations which respect the rapidity ordering on the two-sided plot
of Fig.\ref{fig:two} and \ref{fig:tre}.

Replacing eq.(\ref{fpro}) into eq.(\ref{fklh}), we find it in
agreement with eq.(\ref{mrpt}), thereby proving that the PT amplitudes and
the FKL amplitudes coincide in the high-energy
limit.

The configuration with all the helicities $\nu = -$ is obtained by
replacing the complex conjugates of eq.(\ref{verc}) and (\ref{contra}) into
eq.(\ref{ntree}), which amounts to change $\prod_i p_{i\perp}$ with
$\prod_i p_{i\perp}^*$ in eq.(\ref{fklh}), in agreement with what we noted
after eq.(\ref{mrpt}).

Also the calculation of the FKL amplitude for the other helicity
configurations of eq.(\ref{neg}) is obtained from the one of eq.(\ref{fklh}),
by taking the suitable complex conjugates of the contractions (\ref{contra}),
\begin{eqnarray}
M(p_A,-; p_0,-;...; p_{n+1},+; p_B,+) &=& M(p_A,+; p_0,+;...; p_{n+1},+;
p_B,+)\, ,\nonumber\\
M(p_A,+; p_0,+;...; p_{n+1},-; p_B,-) &=& M(p_A,-; p_0,-;...; p_{n+1},-;
p_B,-)\, ,\label{fklhel}\\ &=& \left({p_{n+1\perp}\over p_{n+1\perp}^*}
\right)^2\, M(p_A,+; p_0,+;...; p_{n+1},+; p_B,+)\, ,\nonumber
\end{eqnarray}
with helicities $\nu_i = +$ and $i=1,...,n$. Eq.(\ref{fklhel}) is in
agreement with eq.(\ref{othel}).

Finally, we note that the two helicities of each gluon
produced along the ladder exchanged in the $\hat t$ channel contribute on
equal footing to the FKL amplitude (\ref{ntree}). Indeed changing the helicity
of gluon $p_i$ we must
take the complex conjugate of eq.(\ref{verc}), and we obtain that the
amplitude (\ref{fklh}) changes only by a phase,
\begin{eqnarray}
& & M(p_A,+; p_0,+;...; p_{j-1},+; p_j,-; p_{j+1},+;...; p_{n+1},+; p_B,+) =
\label{chh}\\
& & {p_{i\perp} q_{i\perp} q_{i+1\perp}^*\over p_{i\perp}^* q_{i\perp}^*
q_{i+1\perp}}\, M(p_A,+; p_0,+;...; p_{n+1},+; p_B,+)\, .\nonumber
\end{eqnarray}

\section{Conclusions}

In this work we have given a unified description of tree-level multigluon
amplitudes in the multiregge kinematics. Representing the color flows in terms
of color lines in the fundamental representation of SU($N_c$), we have shown
that the leading color configurations for the PT amplitudes and for
the FKL amplitudes are the ones whose untwisted color lines
are ordered in rapidity on a two-sided plot. For the helicity configurations
they have in common, we have shown that the PT amplitudes and the FKL
amplitudes are equal, without any phase arbitrariness.

A corollary of the calculation of the FKL amplitude at fixed helicity is that
changing the helicity of a gluon along the gluon ladder changes the
FKL amplitude only by a phase, thus
the two helicities of each gluon emitted along the
ladder contribute equally to an unpolarized production rate
computed from the FKL amplitude.

Finally, as remarked in the Introduction, including the leading logarithmic
contributions, in $\ln(\hat s/\hat t)$, of the loop corrections to
eq.(\ref{ntree}) modifies the propagator of the gluon of momentum $q_i$
exchanged in the $\hat t$ channel by the factor \cite{FKL}
\begin{equation}
{1\over\hat{t}_i} \rightarrow {1\over\hat{t}_i} \left(-{\hat{s}_{i-1,i}
\over\hat{t}_i}\right)^{\alpha(\hat{t}_i)}\, ,\label{sud}
\end{equation}
with $i=1,...,n$ and $\alpha(\hat{t}_i)$ a function of the
loop-momentum integral. Thus the FKL amplitude with the
leading-logarithmic loop corrections retains the ladder structure of
eq.(\ref{ntree}) and the color structure of eq.(\ref{fpro}),
so the dominant color configurations are still
the ones whose untwisted lines are ordered in rapidity on the
two-sided plot. Even though this is a simple observation from the standpoint
of the FKL amplitudes, it is far from being obvious when we consider the
color decomposition of multigluon amplitudes in a helicity basis, since the
color structure of the tree-level amplitudes (\ref{one}) does not
describe all the possible color configurations of $n$ gluons, more
configurations appearing in the color decomposition at the loop level
\cite{bk}.

Thus the invariance of the color structure of the FKL amplitude from the
tree level to the loop level seems to imply that the additional color
configurations that appear in the decomposition of loop-level multigluon
amplitudes in a helicity basis should not give a leading contribution in the
multiregge kinematics.

\section*{Acknowledgements}

I wish to thank Lev Lipatov and Mark W\"usthoff for useful discussions.

\appendix
\section{Multiparton kinematics}

We consider the production of $n+2$ gluons of momentum $p_i$, with
$i=0,...,n+1$ and $n\ge 0$, in the scattering between two gluons of momenta
$p_A$ and $p_B$. Naming the momentum in the beam direction, $p_{_{||}}$,
and in plane transverse to the beam, $p_{\perp}$, and using
light-cone coordinates, $p^{\pm}= p_0\pm p_{_{||}}$, with scalar
product $p\cdot q = (p^+q^- + p^-q^+)/2 - p_{\perp}\cdot q_{\perp}$,
the gluon 4-momenta are,
\begin{eqnarray}
p_A &=& \left(p_A^+, 0; 0, 0\right)\, ,\nonumber \\
p_B &=& \left(0, p_B^-; 0, 0\right)\, ,\label{in}\\
p_i &=& \left(|p_{i\perp}| e^{y_i}, |p_{i\perp}| e^{-y_i};
|p_{i\perp}|\cos{\phi_i}, |p_{i\perp}|\sin{\phi_i}\right)\, ,\nonumber
\end{eqnarray}
where to the left of the semicolon we have the + and -
components, and to the right the transverse components.
$y$ is the gluon rapidity and $\phi$ is the azimuthal angle between the
vector $p_{\perp}$ and an arbitrary vector in the transverse plane.
Throughout the paper we use the complex notation for the transverse momenta,
\begin{equation}
p_{\perp} = |p_{\perp}| e^{i\phi}\, .\label{com}
\end{equation}
{}From the momentum conservation,
\begin{eqnarray}
0 &=& \sum_{i=0}^{n+1} p_{i\perp}\, ,\nonumber \\
p_A^+ &=& \sum_{i=0}^{n+1} |p_{i\perp}| e^{y_i}\, ,\label{nkin}\\
p_B^- &=& \sum_{i=0}^{n+1} |p_{i\perp}| e^{-y_i}\, ,\nonumber
\end{eqnarray}
the Mandelstam invariants may be written as,
\begin{eqnarray}
\hat s &=& 2 p_A\cdot p_B = \sum_{i,j=0}^{n+1} |p_{i\perp}||p_{j\perp}|
e^{y_i-y_j} \nonumber\\ \hat s_{Ai} &=& -2 p_A\cdot p_i = -\sum_{j=0}^{n+1}
|p_{i\perp}||p_{j\perp}| e^{-(y_i-y_j)} \label{inv}\\ \hat s_{Bi}
&=& -2 p_B\cdot p_i = -\sum_{j=0}^{n+1} |p_{i\perp}||p_{j\perp}|
e^{y_i-y_j} \nonumber\\ \hat s_{ij} &=& 2 p_i\cdot p_j = 2 |p_{i\perp}|
|p_{j\perp}| \left[\cosh (y_i-y_j) - \cos (\phi_i-\phi_j) \right]\, .\nonumber
\end{eqnarray}

\section{Massless-spinor algebra}

The spinors (\ref{cpro}) are normalized as\footnote{See ref.\cite{mp} for a
summary of the properties of the spinor algebra.}
\begin{equation}
\langle p\pm| \gamma_{\mu} |p\pm\rangle = 2p_{\mu}\, ,\label{norm}
\end{equation}
and their relative phases are chosen so that under the charge-conjugation
operation $C$
\begin{equation}
|p\mp\rangle = |p\pm\rangle^c = C |p\pm\rangle^*\, ,\label{ket}
\end{equation}
with charge-conjugation matrix satisfying the algebra \cite{bd}
\begin{eqnarray}
\gamma_{\mu} &=& - C \gamma_{\mu}^* C^{-1}\, ,\label{cha}\\
C &=& C^{-1} = C^{\dagger} = C^T\, .\label{chab}
\end{eqnarray}
Eq.(\ref{cha}) determines C up to a phase $C=e^{i\alpha}\gamma_2$, and
the further condition (\ref{chab}) fixes it as $C=\pm i\gamma_2$.
{}From eq.(\ref{cha}) the transposed spinor transforms as,
\begin{equation}
\langle p\mp| =\, ^c\langle p\pm| = -^*\langle p\pm| C\, .\label{bra}
\end{equation}
We use the chiral representation of the $\gamma$-matrices \cite{bd},
\begin{equation}
\gamma^0 = \left( \begin{array}{cc} 0 & I\\ I & 0\end{array} \right),
\qquad \gamma^i = \left( \begin{array}{cc} 0 & -\sigma^i\\ \sigma^i & 0
\end{array} \right)\, ,\label{gam}
\end{equation}
with $I$ the $2\times 2$ unit matrix and $\sigma^i$ the Pauli matrices.
Solving then the Dirac equation $\not p\psi(p) = 0$ and using the normalization
condition (\ref{norm}) and the complex notation (\ref{com}), the spinors
for the gluon momenta (\ref{in}) are,
\begin{equation} \begin{array}{cc}
\psi_+(p) = e^{i\gamma}\, \left( \begin{array}{c} \sqrt{p^+}\\ \sqrt{p^-} e^{i
\phi}\\ 0\\ 0\end{array}\right) & \psi_-(p) = e^{-i\gamma}\, \left(
\begin{array}{c} 0\\ 0\\ \sqrt{p^-} e^{-i\phi}\\ -\sqrt{p^+}\end{array}\right)
\\ \\ \psi_+(p_A) = e^{i\alpha}\, \left( \begin{array}{c} \sqrt{p_A^+}\\ 0\\
0\\ 0\end{array} \right) & \psi_-(p_A) = e^{-i\alpha}\, \left( \begin{array}{c}
0\\ 0\\ 0\\ -\sqrt{p_A^+} \end{array}\right)\\ \\ \psi_+(p_B) = e^{i\beta}\,
\left( \begin{array}{c} 0\\ \sqrt{p_B^-} \\ 0\\ 0\end{array}\right) &
\psi_-(p_B) = e^{-i\beta}\, \left( \begin{array}{c} 0\\ 0\\
\sqrt{p_B^-}\\ 0\end{array}\right) \end{array}\label{spin}
\end{equation}
where the relative sign between $\psi_+(p)$ and $\psi_-(p)$ is fixed by
choosing the charge-conjugation matrix as $C=i\gamma_2$.
The overall phases $e^{i\gamma},\, e^{i\alpha}$ and $e^{i\beta}$ are of
course arbitrary. Without losing generality we fix them to 1. Using
the spinor representation (\ref{spin}), the spinor products for the
momenta (\ref{in}) are
\begin{eqnarray}
\langle p_i p_j\rangle &=& {p_j^+p_{i\perp} - p_i^+p_{j\perp}\over
\sqrt{p_i^+p_j^+}}\, ,\nonumber\\ \langle p_A p_i\rangle &=& -\sqrt{p_A^+
\over p_i^+}\, p_{i\perp}\, ,\label{spro}\\ \langle p_i p_B\rangle &=&
-\sqrt{p_B^-\over p_i^-}\, |p_{i\perp}|\, ,\nonumber\\ \langle p_A p_B\rangle
&=& -\sqrt{\hat s}\, .\nonumber
\end{eqnarray}
Using the spinors (\ref{spin}), the momentum conservation (\ref{nkin}) and
the Mandelstam invariants (\ref{inv}),
it is straightforward to check that the spinor products (\ref{spro})
satisfy the identities,
\begin{eqnarray}
\langle p k\rangle &=& - \langle k p\rangle\, ,\nonumber\\
\langle p k\rangle^* &=& \left[kp\right]\, ,\label{flip}\\
\langle p k\rangle \left[kp\right] &=& 2p\cdot k = |\hat{s}_{pk}|\, ,\nonumber
\end{eqnarray}
which entail that the spinor products may be regarded as the complex square
roots of the Mandelstam invariants.

\section{The gluon polarizations}

For the gluon polarizations we use the representation (\ref{hpol})
\begin {equation}
\epsilon_{\mu}^{\pm}(p,k) = \pm {\langle p\pm |\gamma_{\mu}| k\pm\rangle\over
\sqrt{2} \langle k\mp | p\pm \rangle}\, .\label{hpolap}
\end{equation}
Using the representations (\ref{gam}) for the $\gamma$-matrices and
(\ref{spin}) for the spinors, we obtain
\begin{eqnarray}
\epsilon_{\mu}^+(p_i, p_A) &=& -{p_{i\perp}^*\over p_{i\perp}}\, \left(
{\sqrt{2}\, p_{i\perp}\over p_i^-}, 0; {1\over\sqrt{2}}, {i\over\sqrt{2}}
\right)\, ,\nonumber\\ \epsilon_{\mu}^+(p_B, p_A) &=& - \left(0, 0;
{1\over\sqrt{2}}, {i\over\sqrt{2}}\right) ,\label{exh}\\
\epsilon_{\mu}^+(p_A, p_B) &=& \left(0, 0; {1\over\sqrt{2}}, -{i\over
\sqrt{2}}\right)\, ,\nonumber\\ \epsilon_{\mu}^+(p_i, p_B) &=& \left(0,
{\sqrt{2}\, p_{i\perp}^*\over p_i^+}; {1\over\sqrt{2}}, -{i\over\sqrt{2}}
\right)\, ,\nonumber
\end{eqnarray}
in light-cone coordinates.

The decomposition of a polarization vector in light-cone or Sudakov
components is \cite{lipat},
\begin{eqnarray}
\epsilon_L^{\mu}(p) &=& \epsilon_{L\perp}^{\mu} - {p\cdot\epsilon_{L\perp}\over
p\cdot p_A}\, p_A^{\mu}\, ,\label{lheap}\\
\epsilon_R^{\mu}(p) &=& \epsilon_{R\perp}^{\mu} - {p\cdot\epsilon_{R\perp}\over
p\cdot p_B}\, p_B^{\mu}\, .\nonumber
\end{eqnarray}
The polarizations $\epsilon_L^{\mu}(p),\, \epsilon_R^{\mu}(p)$ and
$\epsilon_{L\perp}^{\mu},\, \epsilon_{R\perp}^{\mu}$ for a momentum $p$ not
in the beam direction are related by a gauge tranformation,
\begin{equation}
\epsilon_R^{\mu}(p) = \epsilon_L^{\mu}(p) + 2\, {\epsilon_{L,\perp}\cdot p\over
|p_{\perp}|^2}\, p^{\mu}\, .\label{gaa}
\end{equation}
Using the complex notation (\ref{com}) the gauge transformation (\ref{gaa})
for the transverse components may be written as,
\begin{equation}
\epsilon_{R\perp}(p) = - {p_{\perp}\over p_{\perp}^*} \epsilon_{L\perp}^*(p)\,
,\label{gac}
\end{equation}
for $p\ne p_A$ or $p_B$. We impose then the following standard polarizations,
\begin{eqnarray}
\epsilon_{L\perp}^{\mu\pm}(p) &=& \left(0, 0; {1\over\sqrt{2}}, \pm {i\over
\sqrt{2}}\right)\, ,\label{staa}\\
\epsilon_{L\perp}^{\mu\pm}(p_B) &=& \epsilon_{R\perp}^{\mu\mp}(p_A) = \left(0,
0; {1\over\sqrt{2}}, \pm {i\over\sqrt{2}}\right)\, .\label{stab}
\end{eqnarray}
We determine $\epsilon_L^{\mu}(p)$ using eq.(\ref{staa}) in the definition
(\ref{lheap}), then we find $\epsilon_R^{\mu}(p)$ using the gauge
transformations (\ref{gaa}) and (\ref{gac}) on $\epsilon_L^{\mu}(p)$.
Finally, $\epsilon_L^{\mu}(p_B)$ and $\epsilon_R^{\mu}(p_A)$ are given by
eq.(\ref{stab}) and the definitions (\ref{lheap}).
Comparing the results with eq.(\ref{exh}),
we obtain the following conversion table among the representations
(\ref{hpolap}) and (\ref{lheap}) of the polarizations
\begin{eqnarray}
\epsilon_{\mu}^+(p_i, p_A) &=& -{p_{i\perp}^*\over p_{i\perp}}\,
\epsilon_{L\mu}^+(p_i)\, ,\nonumber\\ \epsilon_{\mu}^+(p_i, p_B) &=&
-{p_{i\perp}^*\over p_{i\perp}}\, \epsilon_{R\mu}^+(p_i)\, \label{conv}\\
\epsilon_{\mu}^+(p_B, p_A) &=& -\epsilon_{L\mu}^+(p_B)\, ,\nonumber\\
\epsilon_{\mu}^+(p_A, p_B) &=& \epsilon_{R\mu}^+(p_A)\, .\nonumber
\end{eqnarray}


\begin{thebibliography}{99}

\bibitem{lip} L.N.~Lipatov, Yad.~Fiz. {\bf 23}, 642 (1976)
[Sov.~J.~Nucl.~Phys. {\bf 23}, 338 (1976)].

\bibitem{FKL} E.A.~Kuraev, L.N.~Lipatov and V.S.~Fadin, Zh.~Eksp.~Teor.~Fiz.
{\bf 71}, 840 (1976) [Sov.~Phys.~JETP {\bf 44}, 443 (1976)].

\bibitem{BFKL} E.A.~Kuraev, L.N.~Lipatov and
V.S.~Fadin, Zh.~Eksp.~Teor.~Fiz. {\bf 72}, 377 (1977) [Sov.~Phys.~JETP
{\bf 45}, 199 (1977)];\\
Ya.Ya.~Balitsky and L.N.~Lipatov, Yad.~Fiz. {\bf 28} 1597 (1978)
[Sov.~J.~Nucl.~Phys. {\bf 28}, 822 (1978)].

\bibitem{pt} S.J.~Parke and T.~Taylor, Phys.~Rev.~Lett. {\bf 56}, 2459 (1986).

\bibitem{vd} V.~Del~Duca, Phys.~Rev.~D {\bf 48}, 5133 (1993).

\bibitem{bj} G.~Veneziano, Phys.~Lett. {\bf 43B}, 413 (1973); \\
J.D.~Bjorken, Phys.~Rev.~D {\bf 45}, 4077 (1992).

\bibitem{cina} Z.~Xu, D.-H.~Zhang and L.~Chang, Nucl.~Phys. {\bf B291}, 392
(1987).

\bibitem{mp} M.L.~Mangano and S.J.~Parke, Phys.~Rep. {\bf 200}, 301 (1991).

\bibitem{kos} D.A.~Kosower, Nucl.~Phys. {\bf B335}, 23 (1990).

\bibitem{lipat} L.N.~Lipatov, Nucl.~Phys. {\bf B365}, 614 (1991).

\bibitem{mark} J.~Bartels and M.~W\"usthoff, preprint DESY 94-016.

\bibitem{bk} Z.~Bern and D.A.~Kosower, Nucl.~Phys. {\bf B362}, 389 (1991).

\bibitem{bd} J.D.~Bjorken and S.D.~Drell, {\sl Relativistic Quantum Mechanics},
McGraw-Hill, 1964.

\end{thebibliography}
\end{document}